\documentclass[fleqn,twoside]{article}
\usepackage[headings]{espcrc2}

\readRCS
$Id: espcrc2.tex,v 1.2 2004/02/24 11:22:11 spepping Exp $
\usepackage{graphicx}
\usepackage[figuresright]{rotating}

\usepackage{epsfig,bbm,psfrag}
\usepackage{amsmath,amsfonts,amssymb}
\usepackage{citesort}
\usepackage[footnotesize]{caption}
\usepackage{url}

\psfrag{tanb}{\small $\tan{\beta}$}
\psfrag{tan b}{\small $\tan{\beta}$}
\psfrag{log(mN)}{\small $\log(m_N)$}
\psfrag{BR(tau -> 3mu)}{\small BR$(\tau \to 3 \mu)$}
\psfrag{BR(tau -> 3e)}{\small BR$(\tau \to 3 e)$}
\psfrag{BR(mu -> 3e)}{\small BR$(\mu \to 3 e)$}
\psfrag{BR(mu ->3e)}{\small BR$(\mu \to 3 e)$}
\psfrag{BR(tau -> mu + gamma)}{\small BR$(\tau \to \mu \gamma)$}
\psfrag{BR(tau -> e + gamma)}{\small BR$(\tau \to e \gamma)$}
\psfrag{BR(mu -> e + gamma)}{\small BR$(\mu \to e \gamma)$}
\psfrag{mod(theta2)}{\small $|\theta_2|$}
\psfrag{mod(theta1)}{\small $|\theta_1|$}

\title{Lepton Flavour Violation in charged leptons within SUSY-seesaw}

\author{E. Arganda\address[UCM]{ Departamento de F\'{\i }sica At\'omica, Molecular y
Nuclear,
Facultad de Ciencias F\'{\i } sicas, \\
Universidad Complutense de Madrid,
Avda. Complutense s/n,
E-28040 Madrid, Spain},
        M. Herrero\address[UAM]{Departamento de F\'{\i }sica Te\'{o}rica C-XI 
        and Instituto de F\'{\i }sica Te\'{o}rica C-XVI, \\
	Universidad Aut\'{o}noma de Madrid,
	Cantoblanco, E-28049 Madrid, Spain}
	\thanks{Talk given at the 10th
	International Workshop on Tau-Lepton Physics, Tau08, 22-25 September
	2008, Novosibirsk (Russia)},
        J. Portol\'es\address[UV]{IFIC, Universitat de Valencia-CSIC,
	Apt. Correus 22085, E-46071, Valencia,Spain},
	A. Rodriguez-Sanchez\addressmark[UAM]
	and
        A.Teixeira\address[LPT]{Laboratoire de Physique Theorique, UMR 8627,
	Pris-Sud XI, Bat.210, F-91405 Orsay Cedex,France}
}
       
\runtitle{}
\runauthor{E. Arganda, M. Herrero, J. Portoles, A. Rodriguez-Sanchez and A.M. Teixeira}

\begin{document}

\begin{abstract}
In this paper we review our main results for Lepton Flavour Violating (LFV)
semileptonic tau decays and muon-electron conversion in nuclei within the
context of two Constrained SUSY-Seesaw Models, the CMSSM and the NUHM. 
The relevant spectrum is that of the  
Minimal Supersymmetric Standard Model extended by 
three right handed neutrinos, $\nu_{R_i}$ and their corresponding 
SUSY partners, ${\tilde \nu}_{R_i}$, ($i=1,2,3$). We use the seesaw 
mechanism for neutrino mass generation and choose a parameterisation of this
mechanism that allows us to incorporate the neutrino data in our analysis of
LFV processes. In addition to the full one-loop results for the rates of these
processes, we will also review the set of 
simple formulas, valid at large $\tan \beta$, 
which are very
useful to compare with present experimental bounds. The sensitivity to SUSY
and Higgs sectors in these processes will also be discussed.
This is a very short summary of the works in Refs.~\cite{Arganda:2008jj} 
and~\cite{Arganda:2007jw} to which we
refer the reader for more details.
\vspace{1pc}
\end{abstract}

\maketitle

\section{Framework for LFV in charged leptons}

From the present neutrino data on neutrino oscillations, we know 
that Lepton Flavour Violation (LFV) occurs in the neutral lepton sector. 
However, we do not know yet if this LFV occurs in the charged lepton as well.
Even if it occurs, we do not know either if these two violations are related or
not. 
Within the Standard Model with mass less neutrinos there is not LFV. Futhermore,
it is extremely suppressed even with massive neutrinos. 
In contrast, in supersymmetric (SUSY) models with Majorana neutrinos 
LFV can be sizeable. 
In particular, we consider here the spectrum of the Minimal Supersymmetric
Standard Model (MSSM) extended with three rigth handed 
neutrinos, $\nu_{R_i}$, and
their SUSY partners, ${\tilde \nu}_{R_i}$ ($i=1,2,3$), and with the seesaw 
mechanism implemented to
generate the neutrino masses, where it is known that large LFV rates occur.
These
 are induced by the soft SUSY breaking slepton masses and are transmitted to 
 the lepton
sector by means of the Yukawa neutrino couplings, which can be large if the
neutrinos are of Majorana type, and via loops of SUSY
particles. Therefore, in the context we work within of SUSY-seesaw models 
the LFV in both the neutral and the charged lepton are
closely related.

Regarding the seesaw mechanism we use the parameterisation proposed 
in ~\cite{Casas:2001sr}, which is very useful to implement the neutrino data 
into our analysis of LFV. With this parameterisation, the $3 \times 3$ 
Yukawa coupling and Dirac mass matrices are set by
$m_D =\,Y_\nu\,v_2 =\,\sqrt {m_N^{\rm diag}} R \sqrt {m_\nu^{\rm diag}}U^{\dagger}_{\rm
MNS}$, with the $3 \times 3$ orthogonal matrix
$R$ defined by three complex angles $\theta_i$ 
($i=1,2,3$) which represent the additional mixing introduced by the right handed
neutrinos. The other quantities in this formula are  $v_{1(2)}= \,v\,\cos (\sin) \beta$, $v=174$ GeV; 
$m_{\nu}^\mathrm{diag}=\, \mathrm{diag}\,(m_{\nu_1},m_{\nu_2},m_{\nu_3})$ denotes the
three light neutrino masses, and  
$m_N^\mathrm{diag}\,=\, \mathrm{diag}\,(m_{N_1},m_{N_2},m_{N_3})$ the three heavy
ones. $U_{\rm MNS}$ is given by
the three (light) neutrino mixing angles $\theta_{12},\theta_{23}$ and $\theta_{13}$, 
and three phases, $\delta, \phi_1$ and $\phi_2$. With this 
parameterisation it is easy to accommodate
the neutrino data, while leaving room for extra neutrino mixings (from the right
handed sector). It further allows for large
Yukawa couplings $Y_\nu \sim \mathcal{O}(1)$ by
choosing large entries in $m^{\rm diag}_N$ and/or $\theta_i$. 

Here we focus in the particular LFV proccesses: 1) semileptonic $\tau \to
\mu PP$ ($PP = \pi^+ \pi^-, \pi^0 \pi^0, K^+ K^-, K^0 \bar{K}^0$), $\tau \to
\mu P$ ($P =
\pi, \eta, \eta^{\prime}$), $\tau \to
\mu V$ ($V = \rho, \phi$) decays and 2) $\mu-e$ conversion in heavy nuclei. The predictions 
in the following are for two different constrained MSSM-seesaw scenarios, 
with universal and non-universal Higgs soft masses. 
The respective parameters (in addition to the
previous neutrino sector parameters) are: 
1) CMSSM-seesaw: $M_0$, $M_{1/2}$, $A_0$ $\tan \beta$, and sign($\mu$), and 
2) NUHM-seesaw: $M_0$, $M_{1/2}$, $A_0$ $\tan \beta$, sign($\mu$),
$M_{H_1}=M_0(1+\delta_1)^{1/2}$ and
$M_{H_2}=M_0(1+\delta_2)^{1/2}$. 

The predictions presented here include a full one-loop computation 
of the SUSY
diagrams contributing to these LFV processes and
 do not use the Leading
 Logarithmic (LLog) nor the mass insertion approximations.
In the case of semileptonic tau decays we have not included the boxes
which are clearly subdominant, but we have included correspondingly: 
the $\gamma$, Z and Higgs
bosons, $h^0$ and $H^0$, mediated diagrams in $\tau \to \mu PP$, 
and the Z boson and $A^0$ Higgs boson mediated diagrams in $\tau \to \mu P$ .
The hadronisation of quark bilinears in all these semileptonic channels 
is performed within the
chiral framework, using Chiral Perturbation Theory~\cite{Weinberg:1978kz} 
to order and Resonance Chiral
Theory ~\cite{Ecker:1988te} whenever the resonances like the $\rho$, etc., 
play a relevant role.
The predictions for the $\mu-e$ conversion rates include 
the full set of SUSY one-loop contributing diagrams, mediated by $\gamma$, Z
 and Higgs bosons, as well as boxes. In this case we have followed very closely 
the general parameterisation and approximations of 
ref.~\cite{Kuno:1999jp}. 
%%%%%%%%%%%%%%%%%%%%%%%%%%
\section{Results and discussion}
Here we present the predictions for BR($\tau \to \mu PP$) ($PP = \pi^+
\pi^-, \pi^0 \pi^0, K^+ K^-, K^0 \bar{K}^0$), BR($\tau \to \mu P$) ($P =
\pi, \eta, \eta^{\prime}$), BR($\tau \to \mu V$) ($V = \rho, \phi$) and
CR($\mu-e$, Nuclei) within the previously described framework and
compare them with the following experimental bounds: BR$(\tau \to \mu \pi^+
\pi^-) < 4.8 \times 10^{-7}$, BR$(\tau \to \mu K^+
K^-) < 8 \times 10^{-7}$, BR$(\tau \to \mu \pi) < 5.8 \times 10^{-8}$,
BR$(\tau \to \mu \eta) < 5.1 \times 10^{-8}$, BR$(\tau \to \mu
\eta^{\prime}) < 5.3 \times 10^{-8}$, BR$(\tau \to \mu \rho) < 2
\times 10^{-7}$, BR$(\tau \to \mu \phi) < 1.3 \times 10^{-7}$,
CR$(\mu-e, {\rm Au}) < 7 \times 10^{-13}$ and CR$(\mu-e, {\rm Ti}) <
4.3 \times 10^{-12}$.
\begin{figure}[h!]
%\begin{center}
%\begin{tabular}{c}
\psfig{file=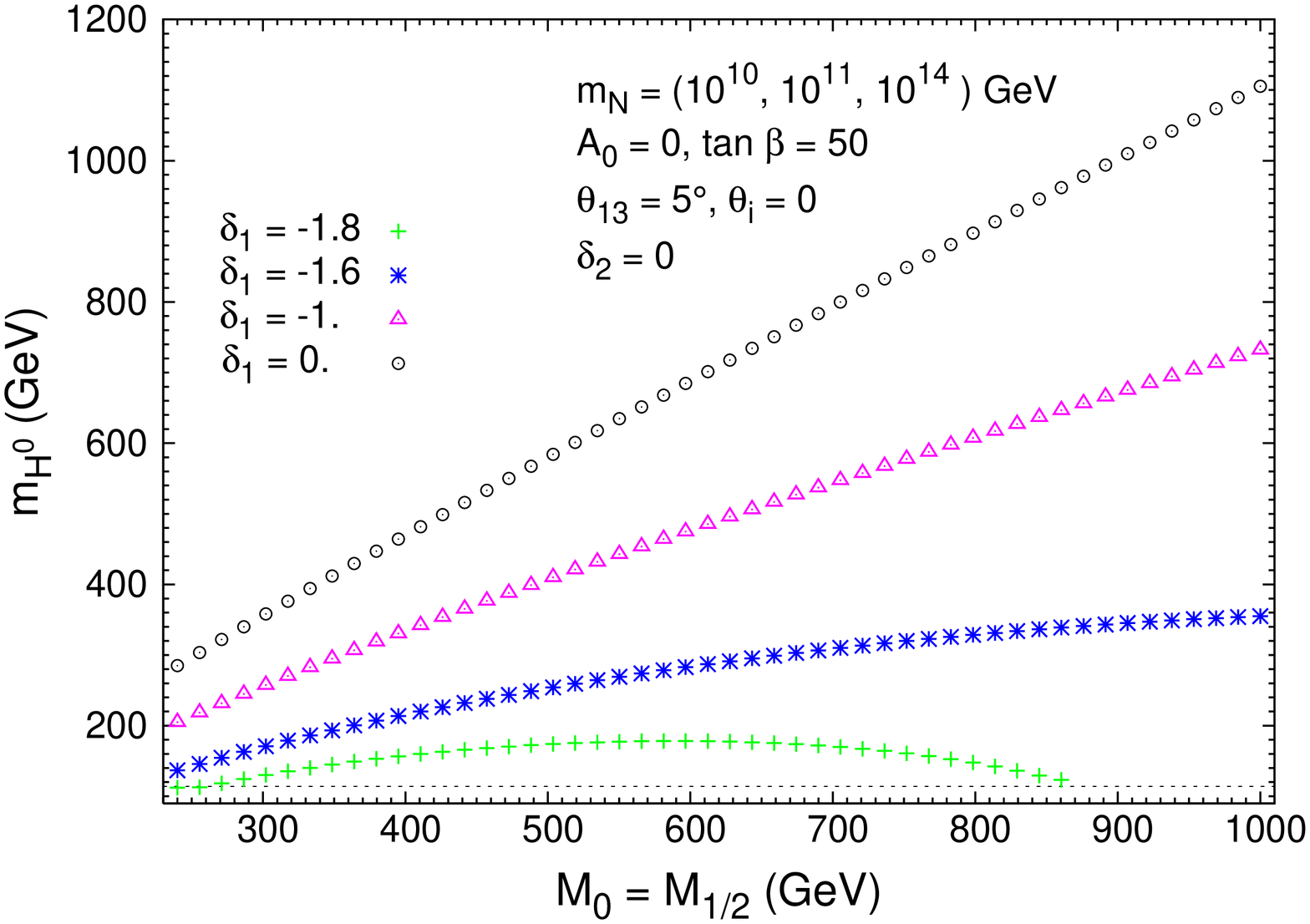,width=50mm,angle=270,clip=}\\
%& \hspace*{1mm}
\psfig{file=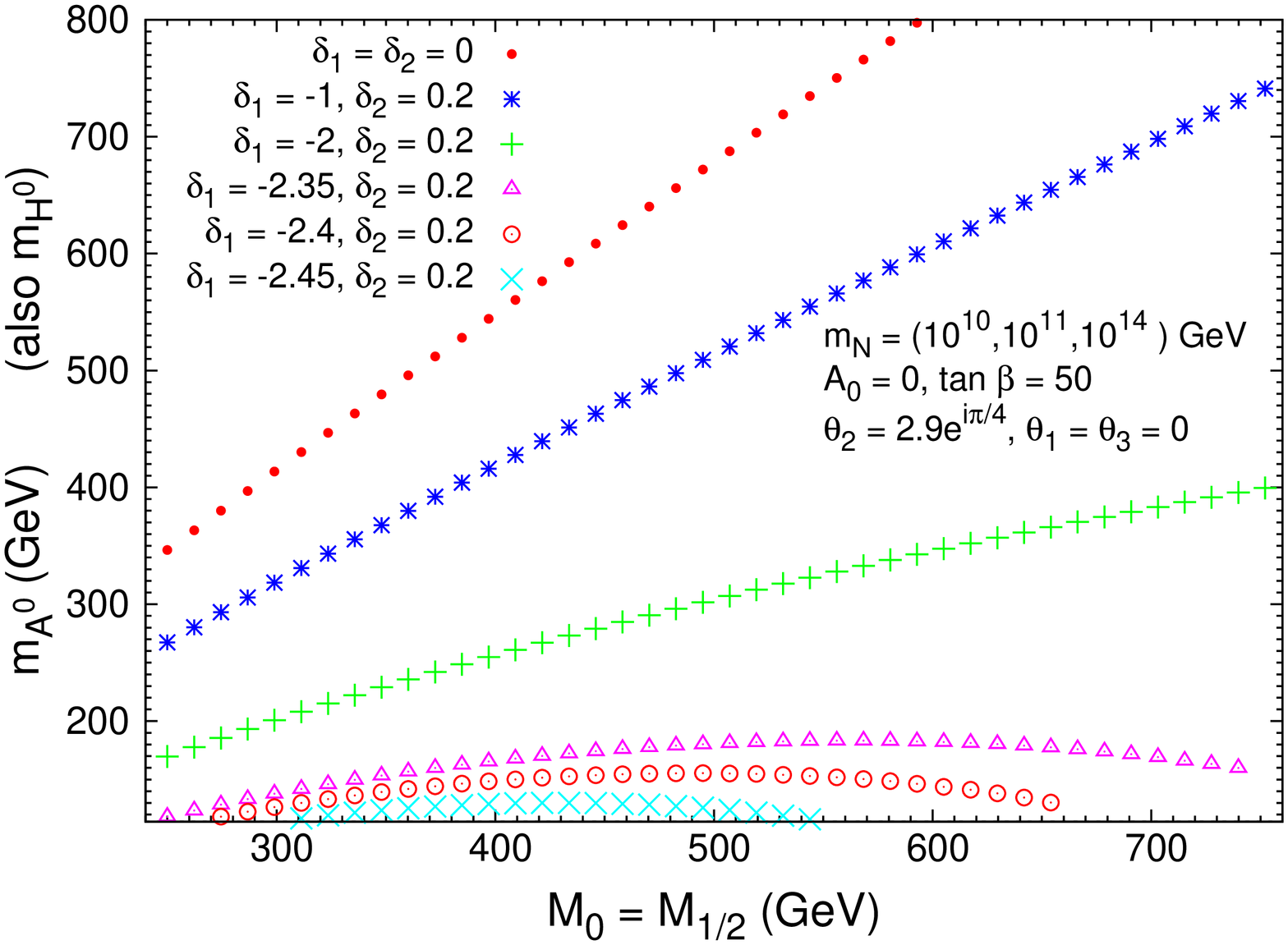,width=50mm,angle=270,clip=}
\caption{Predictions of the Higgs boson masses as a function of
 $M_{\rm SUSY}$ in the CMSSM ($\delta_1 = \delta_2 = 0$) and NUHM 
 ($\delta_1 \neq 0$ and/or $\delta_2 \neq  0$) scenarios.}
\label{fig:1}       % Give a unique label
%\end{tabular}
%\end{center}
\end{figure}
As a general result in LFV processes that can be mediated by Higgs
bosons we have found that the $H^0$ and $A^0$ contributions are relevant at
large $\tan\beta$ if the Higgs masses
are light enough. It is in this aspect where the main difference between the two
considered scenarios lies. Within the CMSSM, light Higgs $H^0$ and $A^0$
bosons are only possible for low $M_{\rm SUSY}$ (here we take $M_{\rm SUSY} =
M_0 = M_{1/2}$ to reduce the number of input parameters). In
contrast, within the NUHM, light Higgs bosons can be obtained even at
large $M_{\rm SUSY}$. In Fig.~\ref{fig:1} it is shown that some
specific choices of $\delta_1$ and $\delta_2$ lead to values of
$m_{H^0}$ and $m_{A^0}$ as low as 110-120 GeV, even for  heavy $M_{\rm SUSY}$
values above 600 GeV. Therefore, the sensitivity to the Higgs sector
is higher in the NUHM.
\begin{figure}[h!]
%\begin{center}
%\begin{tabular}{c}
\psfig{file=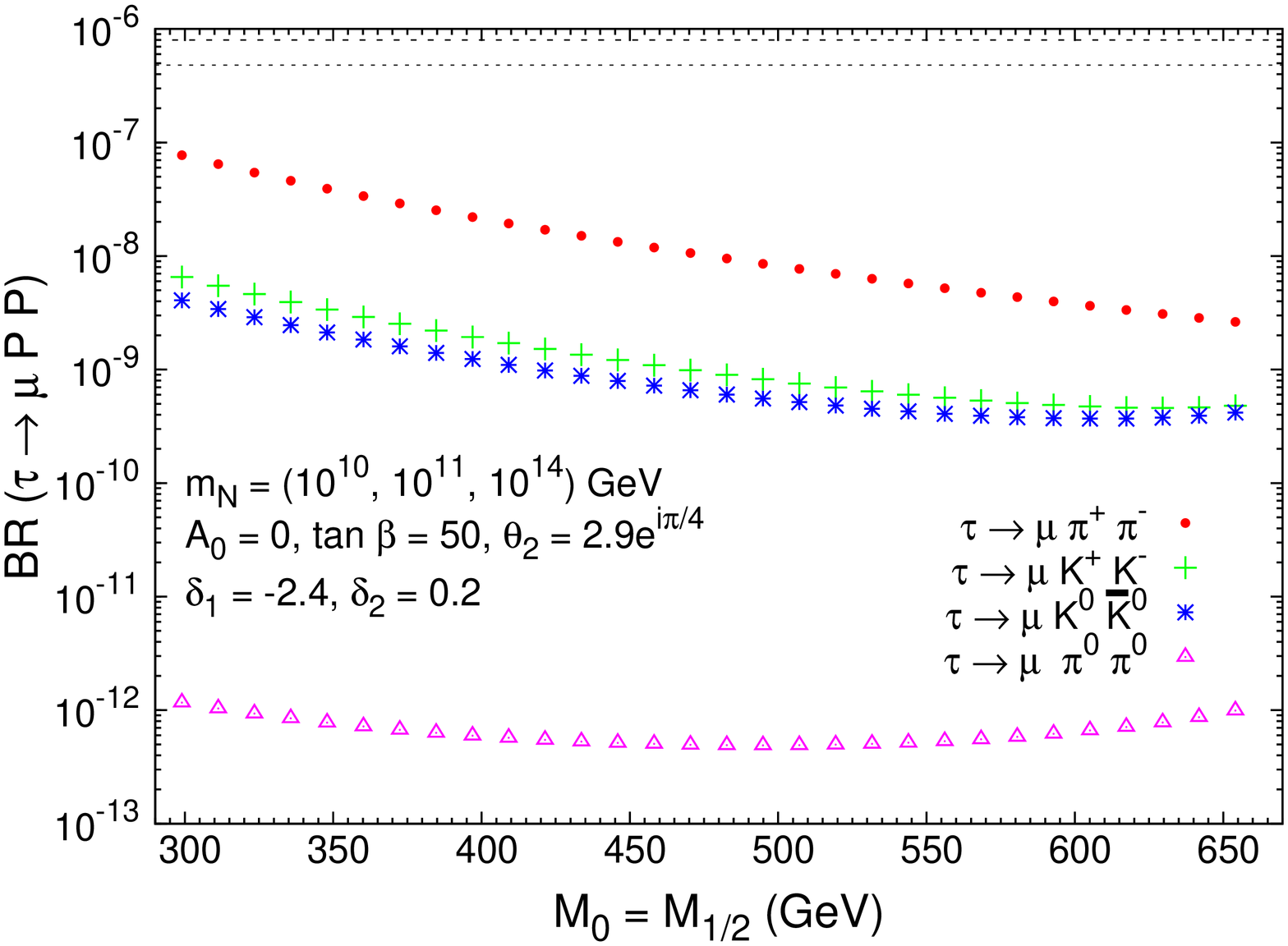,width=50mm,angle=270,clip=}\\
%& \hspace*{1mm}
\psfig{file=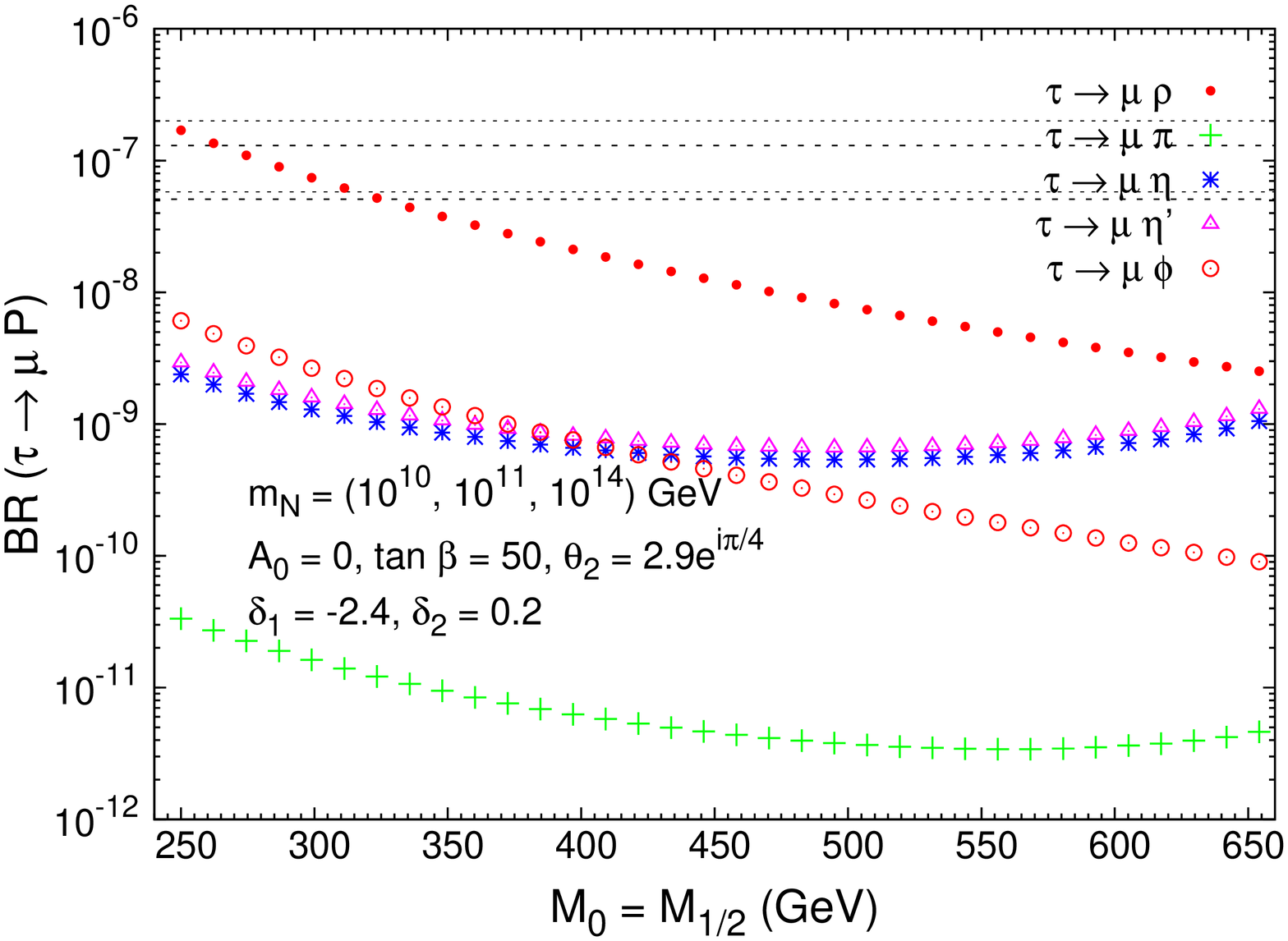,width=50mm,angle=270,clip=}
\caption{Present sensitivity to LFV in semileptonic $\tau$ decays
within the NUHM scenario. The horizontal lines denote
experimental bounds.}
\label{fig:2}       % Give a unique label
%\end{tabular}
%\end{center}
\end{figure}
We start by presenting 
\begin{figure}
%\begin{center}
%\begin{tabular}{c}
\psfig{file=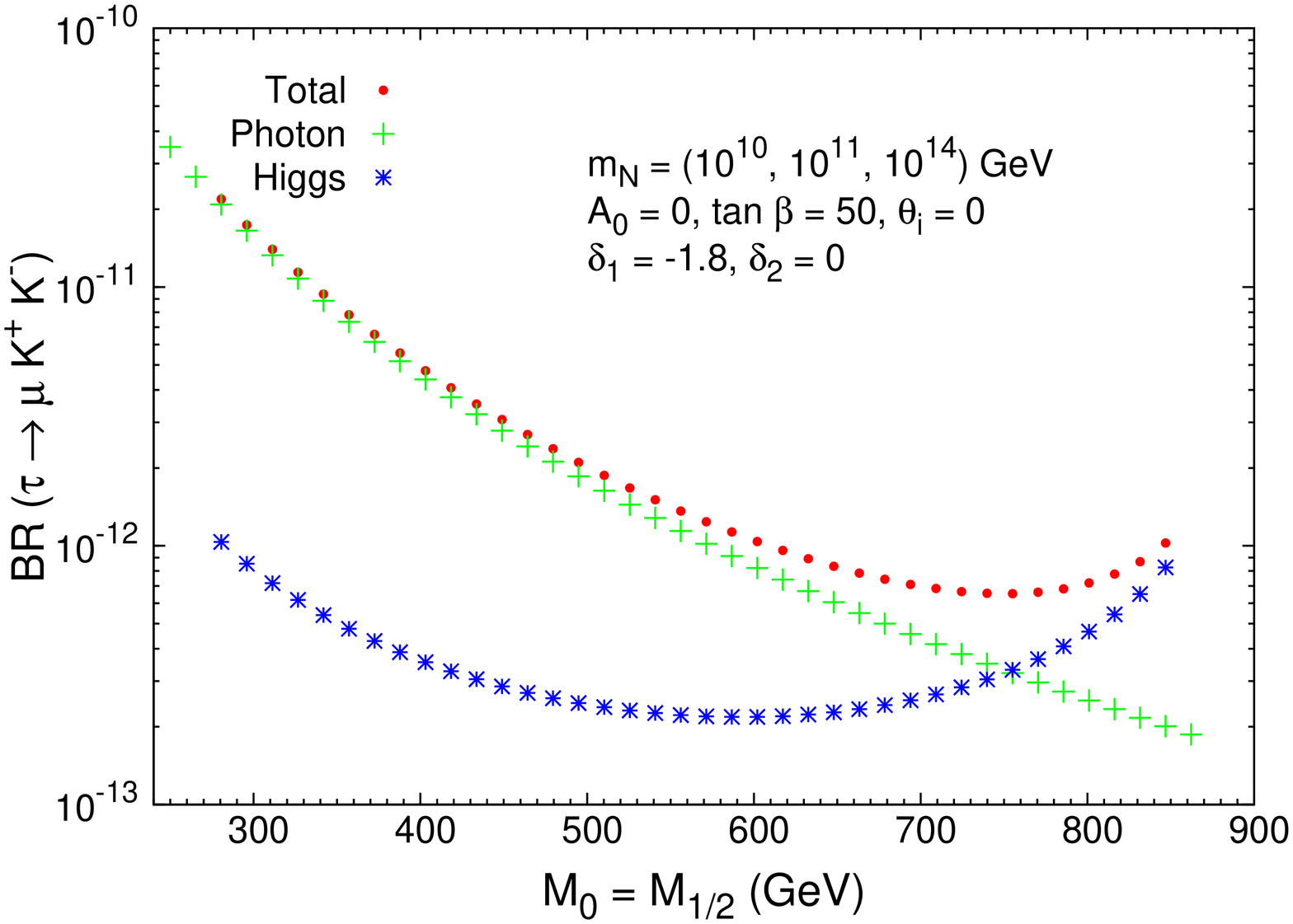,width=50mm,angle=270,clip=}\\
%& \hspace*{1mm}
\psfig{file=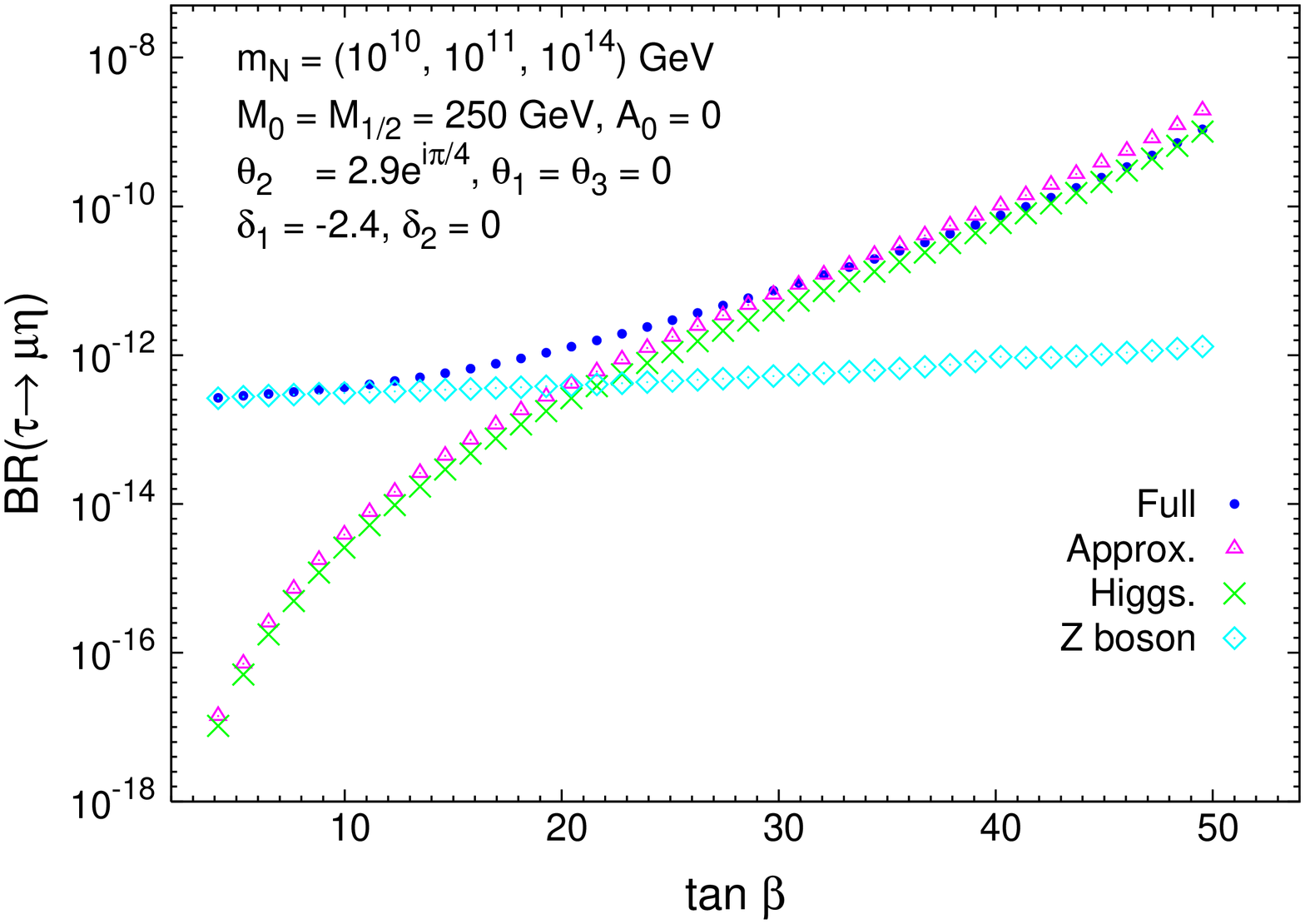,width=50mm,angle=270,clip=}
\caption{Comparison between the various contributions to the semileptonic LFV
tau decays in the NUHM scenario: BR($\tau \to \mu K^+K^-$) plot above, 
BR($\tau \to \mu \eta $) plot below. In this later case, the aproximate result
is also shown for comparison.}
\label{fig:3}       % Give a unique label
%\end{tabular}
%\end{center}
\end{figure}
the results for the semileptonic tau
decays. The mentioned sensitivity to the Higgs sector within the NUHM
scenario can be seen
in Fig.~\ref{fig:2}. Concretely, the BRs of the channels $\tau \to \mu K^+ K^-$,
$\tau \to \mu K^0 \bar K^0$, $\tau \to \mu \pi^0 \pi^0$, $\tau
\to \mu \pi$, $\tau \to \mu \eta$ and $\tau \to \mu \eta^{\prime}$
present a growing behaviour with $M_{\rm SUSY}$, in the large $M_{\rm
SUSY}$ region, due to the contribution of light Higgs bosons,
which is non-decoupling. The decays involving Kaons and $\eta$
mesons are particularly sensitive to the Higgs contributions because
of their strange quark content, which has a stronger coupling to the Higgs bosons.
On the other hand, the largest predicted
rates are for $\tau \to \mu \pi^+ \pi^-$ and $\tau \to \mu \rho$,
dominated by the photon contribution, which are indeed at the present experimental
reach in the low $M_{\rm SUSY}$ region.

\begin{figure}
%\begin{center}
%\begin{tabular}{cc}
\psfig{file=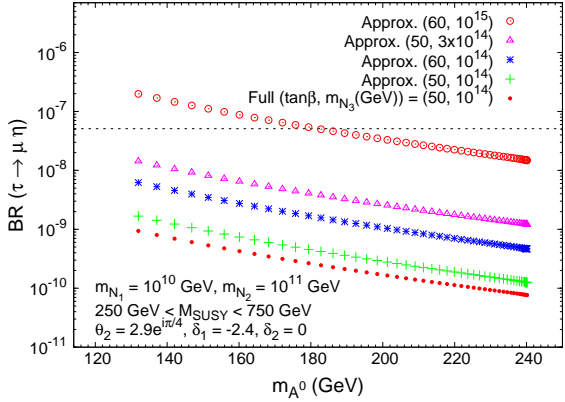,width=55mm,angle=270,clip=}
%& \hspace*{1mm}
\caption{Sensitivity to Higgs sector in $\tau \to \mu \eta$ decays.}
\label{fig:4}       % Give a unique label
%\end{tabular}
%\end{center}
\end{figure}
The comparison between the various contributions to the semileptonic LFV
tau decays in the NUHM scenario for BR($\tau \to \mu K^+K^-$) and
BR($\tau \to \mu \eta $) can be seen in Fig.~\ref{fig:3}. 
It is clear from this
figure, that $\tau \to \mu K^+K^-$ is dominated by the photon contribution,
except in the large $M_{\rm SUSY}$ region , say $M_{\rm SUSY}> 750$ GeV,
and large $\tan \beta$ region, say $\tan \beta \geq 50$,  where the Higgs
boson contribution plays an important role. Similar results are found for 
$\tau \to \mu K^0 \bar{K}^0$. In contrast, in $\tau \to \mu \pi^+\pi^-$, 
the photon contribution dominantes largely the rates in all the studied region
of the parameter space and therefor it is not sensitive at all to the Higgs
sector. In the $\tau \to \mu \pi^0\pi^0$ channel, only the Higgs boson
contributes, but the rates are extremely small. They are indeed much smaller
that decays into kaons due to the fact that 
the Higgs couplings to the pions are proportional to $m_\pi^2$ whereas the the Higgs couplings to the pions 
are proportional to $m_K^2$. On the other hand, 
the $\tau \to \mu \eta $ channel is 
dominated by the $A^0$ Higgs boson contribution for all $M_{\rm SUSY}$ values, and 
for moderate and large $\tan\beta$, say $\tan\beta > 22$. Notice that for 
smaller values 
of $\tan\beta$ it is, however, dominated by the Z boson contribution.  

A set of useful formulae for all these channels, within the mass
insertion approximation which are
valid at large $\tan\beta$, have also being derived by us in~\cite{Arganda:2008jj}.
We include the most relevant of these aproximate formulas here,
for completeness. The approximate results for the $H^0$-mediated contributions 
and 
the $\gamma$-mediated contributions are shown separately for comparison,  
\begin{eqnarray}
{\rm BR}(\tau \to \mu \eta)_{H_{\rm
approx}}{\,\,\,\,\,\,\,\,\,\,\,\,\,\,\,\,\,} =
{\,\,\,\,\,\,\,\,\,\,\,\,\,\,\,\,\,\,\,\,\,\,\,\,\,\,
\,\,\,\,\,\,\,\,\,\,\,\,\,\,\,\,\,\,\,\,\,\,\,\,\,\,\,\,\,\,\,\,\,\,\,\,\,\,\,} &  \nonumber \\
 1.2 \times 10^{-7} \left| \delta_{32} \right|^2 \left( \frac{100}
{m_{A^0}({\rm GeV})}
\right)^4 \left( \frac{\tan \beta}{60} \right)^6 & \nonumber \\ 
%\sim { 
%\frac{1}{7} \times  {\rm BR}_{\rm Sher}\hspace{0.2cm} {\rm PRD}66(2002)57301} &&\nonumber \\
{\rm BR}(\tau \to \mu \eta')_{H_{\rm approx}} {\,\,\,\,\,\,\,\,\,\,\,\,\,\,\,\,} = 
{\,\,\,\,\,\,\,\,\,\,\,\,\,\,\,\,\,\,\,\,\,\,\,\,\,\,
\,\,\,\,\,\,\,\,\,\,\,\,\,\,\,\,\,\,\,\,\,\,\,\,\,\,\,\,\,\,\,\,\,\,\,\,\,\,\,}
& \nonumber \\
1.5 \times 10^{-7} \left| \delta_{32} \right|^2 \left( \frac{100}
{m_{A^0}({\rm GeV})}
\right)^4 \left( \frac{\tan \beta}{60} \right)^6  & \nonumber \\  
%\sim {
%100 \times  {\rm BR}_{\rm Brignole-Rossi}{\rm NPB}701(04)3} \nonumber \\
{\rm BR}(\tau \to \mu \pi)_{H_{\rm approx}} {\,\,\,\,\,\,\,\,\,\,\,\,\,\,\,\,\,} =
{\,\,\,\,\,\,\,\,\,\,\,\,\,\,\,\,\,\,\,\,\,\,\,\,\,\,
\,\,\,\,\,\,\,\,\,\,\,\,\,\,\,\,\,\,\,\,\,\,\,\,\,\,\,\,\,\,\,\,\,\,\,\,\,\,\,}
& \nonumber \\
3.6\times 10^{-10} \left| \delta_{32} \right|^2 \left( \frac{100}
{m_{A^0}({\rm GeV})}
\right)^4 \left( \frac{\tan \beta}{60} \right)^6  & \nonumber \\   
%{\sim {\rm BR}_{\rm Brignole-Rossi}} \nonumber \\ 
{\rm BR}(\tau \to \mu \pi^0 \pi^0)_{H_{\rm approx}} {\,\,\,\,\,\,\,\,} =
{\,\,\,\,\,\,\,\,\,\,\,\,\,\,\,\,\,\,\,\,\,\,\,\,\,\,
\,\,\,\,\,\,\,\,\,\,\,\,\,\,\,\,\,\,\,\,\,\,\,\,\,\,\,\,\,\,\,\,\,\,\,\,\,\,\,}
& \nonumber \\ 
1.3 \times 10^{-10}\left| \delta_{32} \right|^2 \left( \frac{100}
{m_{H^0}({\rm GeV})} \right)^4 \left(
\frac{\tan \beta}{60} \right)^6  & \nonumber \\  
{\rm BR}(\tau \to \mu \pi^+ \pi^-)_{H_{\rm approx}} {\,\,\,\,} =
{\,\,\,\,\,\,\,\,\,\,\,\,\,\,\,\,\,\,\,\,\,\,\,\,\,\,
\,\,\,\,\,\,\,\,\,\,\,\,\,\,\,\,\,\,\,\,\,\,\,\,\,\,\,\,\,\,\,\,\,\,\,\,\,\,\,}
& \nonumber \\
2.6 \times 10^{-10}\left| \delta_{32} \right|^2 \left( \frac{100}
{m_{H^0}({\rm GeV})} \right)^4 \left(
\frac{\tan \beta}{60} \right)^6 & \nonumber \\
{\rm BR}(\tau \to \mu K^+K^-)_{H_{\rm approx}}  =
{\,\,\,\,\,\,\,\,\,\,\,\,\,\,\,\,\,\,\,\,\,\,\,\,\,\,
\,\,\,\,\,\,\,\,\,\,\,\,\,\,\,\,\,\,\,\,\,\,\,\,\,\,\,\,\,\,\,\,\,\,\,\,\,\,\,}
& \nonumber \\
2.8 \times 10^{-8}\left| \delta_{32} \right|^2 \left( \frac{100}
{m_{H^0}({\rm GeV})} \right)^4 \left(
\frac{\tan \beta}{60} \right)^6  & \nonumber \\   
%{\sim \frac{1}{50}
%\times {\rm BR}_{\rm Chen-Geng} \hspace{0.2cm} {\rm PRD}74(2006)35010} \nonumber \\
{\rm BR}(\tau \to \mu K^0{\bar K}^0)_{H_{\rm approx}}  =
{\,\,\,\,\,\,\,\,\,\,\,\,\,\,\,\,\,\,\,\,\,\,\,\,\,\,
\,\,\,\,\,\,\,\,\,\,\,\,\,\,\,\,\,\,\,\,\,\,\,\,\,\,\,\,\,\,\,\,\,\,\,\,\,\,\,}
&  \nonumber \\
3.0 \times 10^{-8}\left| \delta_{32} \right|^2 \left( \frac{100}
{m_{H^0}({\rm GeV})} \right)^4 \left(
\frac{\tan \beta}{60} \right)^6 & \nonumber \\
{\rm BR}(\tau \to \mu \pi^+ \pi^-)_{\gamma_{\rm approx}} {\,\,\,\,}=
{\,\,\,\,\,\,\,\,\,\,\,\,\,\,\,\,\,\,\,\,\,\,\,\,\,\,
\,\,\,\,\,\,\,\,\,\,\,\,\,\,\,\,\,\,\,\,\,\,\,\,\,\,\,\,\,\,\,\,\,\,\,\,\,\,\,}
& \nonumber \\
3.7 \times 10^{-5}  
\left| \delta_{32} \right|^2 \left( \frac{100}
{M_{\rm SUSY}({\rm GeV})}\right)^4 \left( \frac{\tan \beta}{60} \right)^2 & \nonumber \\
%\hspace{0.5cm}
%{{\bf {\rm dominant\,\,\,for\,\,\,all}\,\,\, M_{\rm SUSY}}} \nonumber \\
{\rm BR}(\tau \to \mu K^+ K^-)_{\gamma_{\rm approx}} = 
{\,\,\,\,\,\,\,\,\,\,\,\,\,\,\,\,\,\,\,\,\,\,\,\,\,\,
\,\,\,\,\,\,\,\,\,\,\,\,\,\,\,\,\,\,\,\,\,\,\,\,\,\,\,\,\,\,\,\,\,\,\,\,\,\,\,}
& \nonumber \\
3.0 \times 10^{-6}  
\left| \delta_{32} \right|^2 \left( \frac{100}
{M_{\rm SUSY}({\rm GeV})}\right)^4 \left( \frac{\tan \beta}{60} \right)^2 & \nonumber \\
%{{\bf {\rm dominant \,\,\,if}\,\,\, M_{\rm SUSY}\leq 300\,\, {\rm GeV}}}\nonumber \\
{\rm BR}(\tau \to \mu K^0 \bar{K}^0)_{\gamma_{\rm approx}} {\,\,\,\,}= 
{\,\,\,\,\,\,\,\,\,\,\,\,\,\,\,\,\,\,\,\,\,\,\,\,\,\,
\,\,\,\,\,\,\,\,\,\,\,\,\,\,\,\,\,\,\,\,\,\,\,\,\,\,\,\,\,\,\,\,\,\,\,\,\,\,\,}
& \nonumber \\
1.8\times 10^{-6}  
\left| \delta_{32} \right|^2 \left( \frac{100}
{M_{\rm SUSY}({\rm GeV})}\right)^4 \left( \frac{\tan \beta}{60} \right)^2 & \nonumber \\
%{{\bf {\rm dominant \,\,\,if}\,\,\, M_{\rm SUSY}\leq 250\,\, {\rm GeV}}}\nonumber \\
%{{\rm Compare \,\,\, to}}\,\,{\rm BR}(\tau \to \mu \gamma)_{\rm approx}& = &
% 1.5 \times 10^{-2}
% \left| \delta_{32} \right|^2 \left( \frac{100}
%{M_{\rm SUSY}({\rm GeV})}\right)^4 \left( \frac{\tan \beta}{60}
% \right)^2 {}{>{\rm semil\,\,\, if} 
% M_{\rm SUSY}<1500{\rm GeV}}
%}
\end{eqnarray}

We have shown that the predictions with these formulae agree with the
full results within a factor of about 2. In the case of $\tau \to \mu
\eta$ this comparison is shown in
Figs.~\ref{fig:3} and ~\ref{fig:4}. It is also clear, from Fig.~\ref{fig:3}
that the approximation works much better in the large $\tan \beta$ region,
 $\tan \beta> 22$ where the $H^0$ boson dominates. Similar conclusions are found for $\tau \to \mu
\eta^{\prime}$. The next relevant channel in sensitivity to the Higgs
sector is $\tau \to \mu K^+ K^-$, but
it is still below the present experimental bound. To our knowledge,
there are not experimental bounds yet  available for $\tau \to \mu K^0
\bar K^0$ and $\tau \to \mu \pi^0 \pi^0$.

Finally, the maximum sensitivity to the Higgs sector is found for 
$\tau \to \mu \eta$ and $\tau \to \mu \eta^{\prime}$ channels, 
largely dominated by
the $A^0$ boson exchange. Fig.~\ref{fig:4} shows that BR($\tau \to \mu
\eta$) reaches the experimental bound for large heaviest neutrino mass,
large $\tan\beta$, large $\theta_i$ angles and low $m_{A^0}$. For the choice of input
parameters in this figure, it occurs at 
$m_{N_3} = 10^{15}$ GeV, $\tan\beta = 60$, $\theta_2 = 2.9
e^{i\pi/4}$ and $m_{A^0} = 180$ GeV.

\begin{figure}[h!]
%\begin{center}
%\begin{tabular}{cc}
\psfig{file=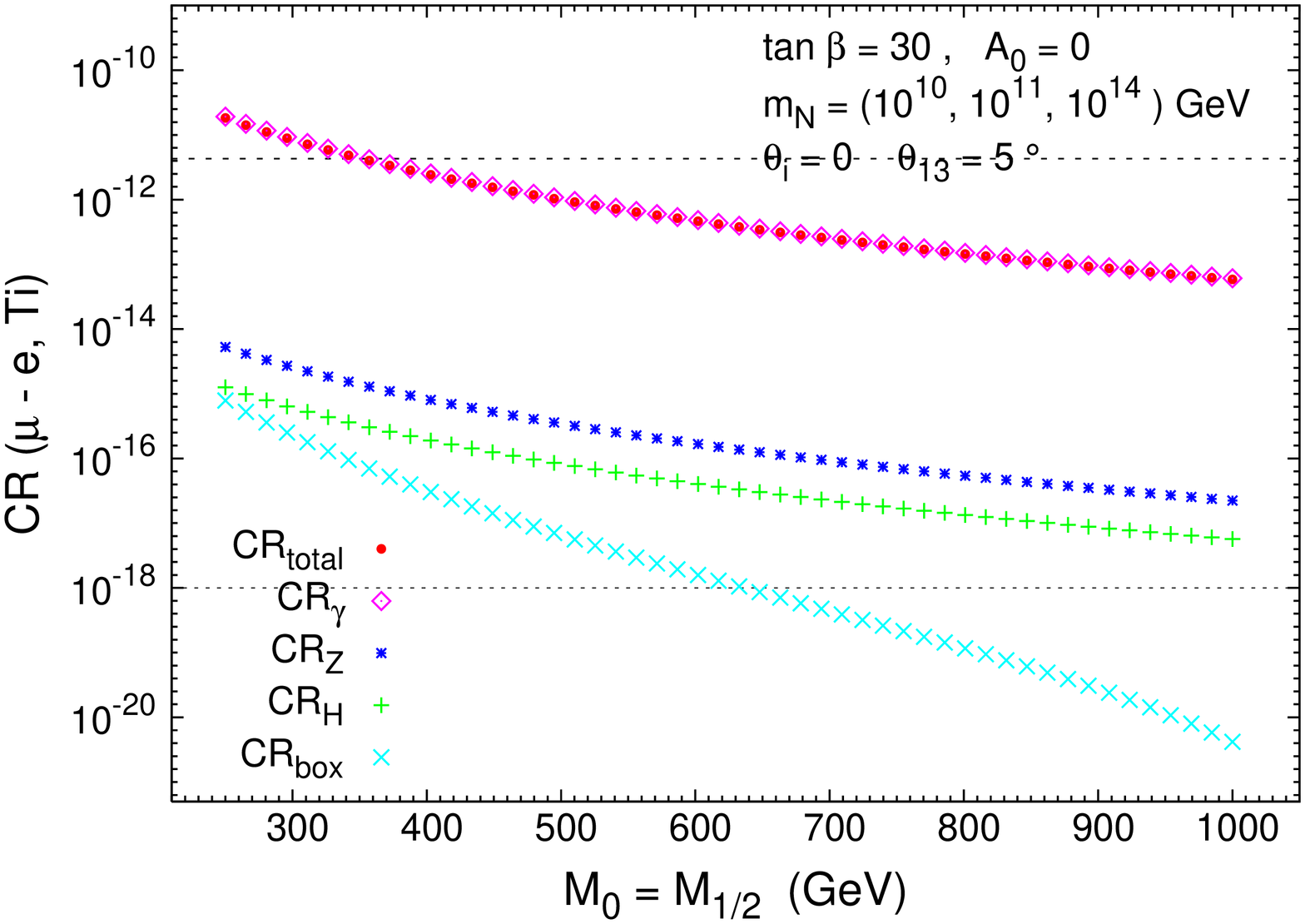,width=50mm,angle=270,clip=}\\
%& \hspace*{1mm}
\psfig{file=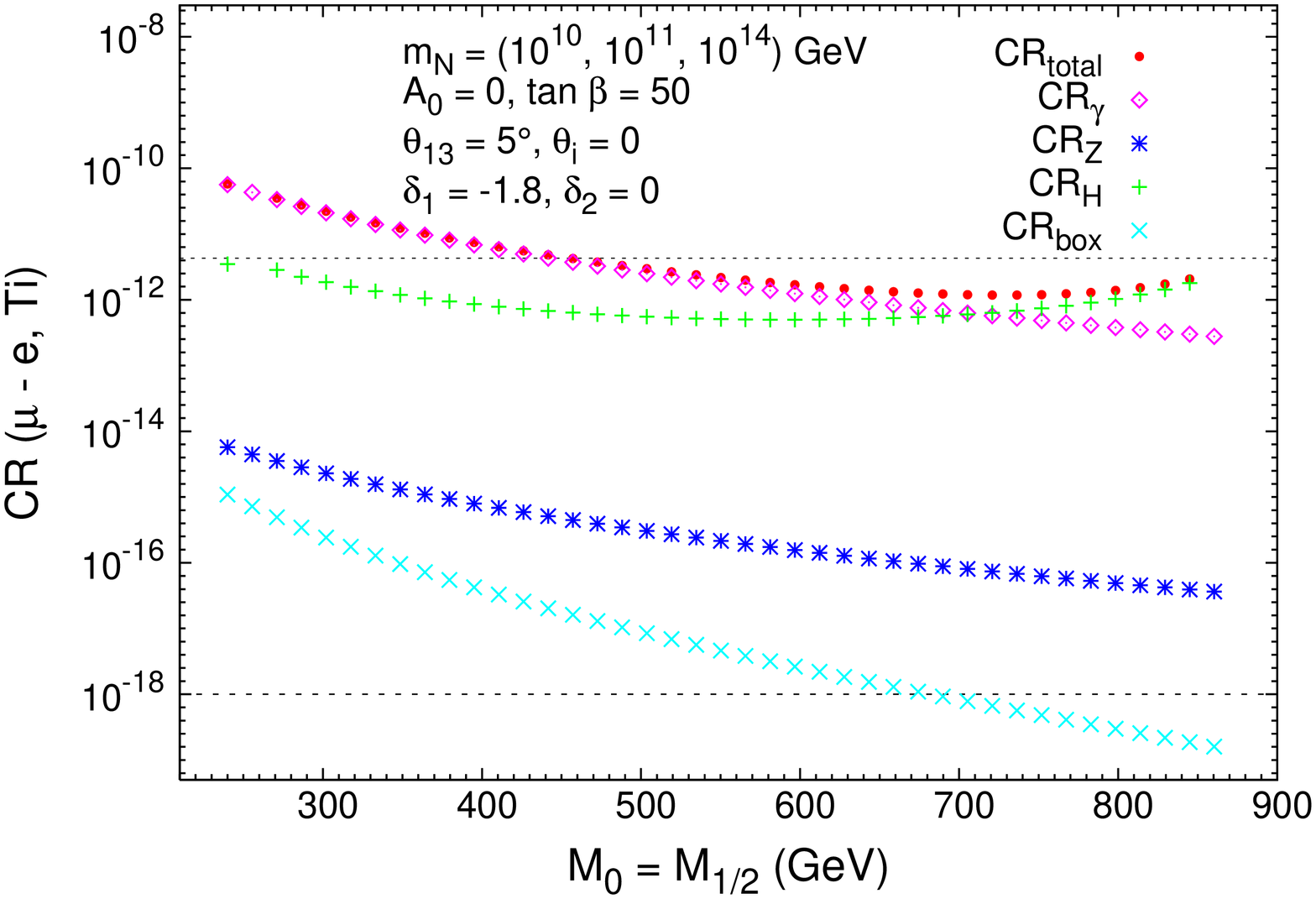,width=50mm,angle=270,clip=}
\caption{Predictions of CR($\mu-e$, Ti) as a function of
 $M_{\rm SUSY}$ in the CMSSM (above) and NUHM (below) scenarios.}
\label{fig:5}       % Give a unique label
%\end{tabular}
%\end{center}
\end{figure}

Next we comment on the results for $\mu-e$ conversion in
nuclei. Fig.~\ref{fig:5} shows our predictions of the conversion rates
for Titanium as a function of $M_{\rm SUSY}$ in both CMSSM and NUHM
scenarios. As in the case of semileptonic tau decays, the sensitivity
to the Higgs contribution is only manifest in the NUHM scenario. The
predictions for CR($\mu-e$, Ti) within the CMSSM scenario are largely
dominated by the photon contribution and present a decoupling
behaviour at large $M_{\rm SUSY}$. In this case the present
experimental bound is only reached at low $M_{\rm SUSY}$. The
perspectives for the future are much more promising. If the announced
sensitivity by PRISM/PRIME of $10^{-18}$ is finally attained, the full studied
range of $M_{\rm SUSY}$ will be covered. 

Fig.~\ref{fig:5} also illustrates that within the NUHM scenario the
Higgs contribution dominates at large $M_{\rm SUSY}$ for light Higgs
bosons. The predicted rates are close to the present experimental bound
not only in the low $M_{\rm SUSY}$ region but also for heavy SUSY
spectra. As in the previous semileptonic tau decays, we have also found 
a simple formula for the conversion rates, within the mass
insertion approximation, which is
valid at large $\tan\beta$~\cite{Arganda:2007jw} and can be used for
further analysis. This is dominated by the Higgs $H^0$ contribution and is given
by,
\begin{eqnarray}
{\rm CR} (\mu- e, {\rm Nucleus})|_{H{\rm approx}} {\,\,\,\,\,\,\,\,\,\,\,\,\,\,\,\,\,\,\,\,\,\,\,\,\,\,\,} \simeq & {\,\,\,\,\,\,\,\,\,}\nonumber \\
\frac{m_\mu^5\, G_F^2\, \alpha^3 \,Z_{\rm eff}^4\, 
F_p^2}{8 \pi^2 \,Z}\, (Z + N)^2
 \left|g_{LS}^{(0)} \right|^2 \, 
\frac{1}{\Gamma_{\rm capt}}\,,& \nonumber \\
g_{LS}^{(0)} =\frac{g^2}{48 \pi^2}G_S^{(s,p)}\frac{m_\mu m_s}{m^2_{H^0}}
\delta_{21} (\tan\beta)^3 & \nonumber \\
\end{eqnarray}   
It shows clearly the relevant features: the $\tan^6\beta$ enhacement of the 
rates, the Higgs mass
dependence, $\propto 1/{m^4_{H^0}}$, and the strange quark mass dependence,
$\propto m_s^2$. 

\begin{figure}[h!]
%\begin{center}
%\begin{tabular}{cc}
\psfig{file=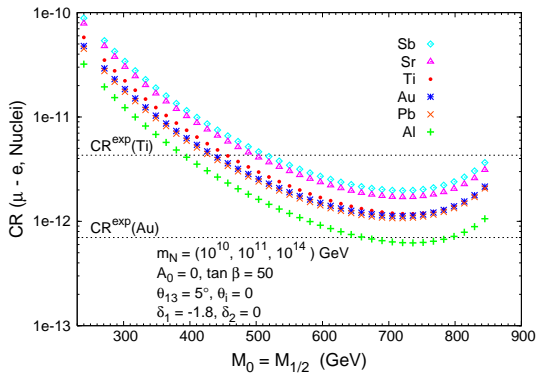,width=50mm,angle=270,clip=}
%& \hspace*{1mm}
\caption{Present sensitivity to LFV in $\mu-e$ conversion for several
nuclei within NUHM.}
\label{fig:6}       % Give a unique label
%\end{tabular}
%\end{center}
\end{figure}

The predictions of the $\mu-e$ conversion rates for several nuclei are
collected in Fig.~\ref{fig:6}. We can see again the growing behaviour
with $M_{\rm SUSY}$ in the large $M_{\rm SUSY}$ region due to the
non-decoupling of the Higgs contributions. At present, the most
competitive nucleus for LFV searches is Au where, for the choice of
input parameters in this figure, all the predicted rates are above the
experimental bound. We have also shown in~\cite{Arganda:2007jw} that $\mu-e$ conversion in
nuclei is extremely sensitive to $\theta_{13}$, similarly to $\mu \to
e \gamma$ and $\mu \to 3e$ and, therefore, a future
measurement of this mixing angle can help in the searches of LFV in
the $\mu-e$ sector.

In conclusion, we have shown that semileptonic tau decays nicely complement
the searches for LFV in the $\tau-\mu$ sector, in addition to
$\tau \to \mu \gamma$. The future prospects for $\mu-e$ conversion in
Ti are the most promising for LFV searches. Both processes,
semileptonic tau decays and $\mu-e$ conversion in nuclei are indeed more
sensitive to the Higgs sector than $\tau \to 3 \mu$.

\section*{Acknowledgements}

M.J. Herrero would like to thank the organizers of this tau-08 conference 
for the invitation to participate
in this interesting and fruitful event. She also acknowledges project
FPA2006-05423 of Spanish MEC for finantial support.

\end{document}